\documentclass[aps,prc,preprint,superscriptaddress,showpacs,12pt ]{revtex4}

\usepackage{epsfig}

\usepackage{graphicx}

\def\half{{\textstyle{1\over 2}}}

\def\nn{\nonumber}

\newcommand{\beq}{\begin{equation}}

\newcommand{\eeq}{\end{equation}}

\newcommand{\beqa}{\begin{eqnarray}}

\newcommand{\eeqa}{\end{eqnarray}}

\newcommand{\bma}{\begin{mathematica}}

\newcommand{\ema}{\end{mathematica}}

\newcommand{\bm}{\mbox {\boldmath}}

\usepackage{epsfig}

\begin{document}

\title{Study of $\eta$ photoproduction on the proton in a chiral constituent
quark approach via one-gluon-exchange model}

\author{Jun He}

\email[]{jun.he@cea.fr} \affiliation{Institut de Recherche sur les
lois Fondamentales de l'Univers, DSM/IRFU, CEA/Saclay, 91191
Gif-sur-Yvette, France}

\author{B. Saghai}

\email[]{bijan.saghai@cea.fr} \affiliation{Institut de Recherche sur
les lois Fondamentales de l'Univers, DSM/IRFU, CEA/Saclay, 91191
Gif-sur-Yvette, France}

\author{Zhenping Li}

\affiliation{Department of Computer and Information Science,
University of   Maryland, MD 20783, USA}

\date{\today}

\begin{abstract}

A formalism based on a chiral quark model ($\chi$QM) approach
complemented with a one-gluon exchange model, to take into account
the breakdown of the $SU(6)\otimes O(3)$ symmetry, is presented. The
configuration mixing of wave functions for nucleon and resonances
are derived.
With few adjustable parameters, differential cross-section and
polarized beam asymmetry for the $\gamma~p~\to~\eta~p$ process are
calculated and successfully compared with the data in the
centre-of-mass energy range from threshold up to 2 GeV. The known
resonances $S_{11}(1535)$, $S_{11}(1650)$, $P_{13}(1720)$,
$D_{13}(1520)$, and $F_{15}(1680)$, as well as two new $S_{11}$ and
$D_{15}$ resonances are found to be dominant in the reaction
mechanism. Besides, connections among the scattering amplitudes of
the $\chi$QM approach and the helicity amplitudes, as well as decay
widths of resonances are established. Possible contributions from
the so-called "missing resonances" are investigated and found to be
negligible.

\end{abstract}

\pacs{  13.60.Le, 12.39.Fe, 12.39.Jh,  14.20.Gk }






\maketitle

\section{Introduction}
\label{Sec:Intro}

Electromagnetic production of mesons on the nucleon offers a great
opportunity to deepen our understanding of the baryon resonances
properties. In recent years, intensive experimental efforts have
been devoted to the measurement of observables for the processes of
pseudoscalar and vector mesons production, using electron and/or
photon beam facilities.

In the present work we investigate the reaction
$\gamma~p~\to~\eta~p$, in the range of centre-of-mass total energy
from threshold up to $W~\approx$ 2 GeV, in order to interpret a
large amount of high quality data released from various facilities,
namely, differential cross-section data by the following
collaborations: MAMI~\cite{Krusche1995}, CLAS~\cite{Dugger2002},
CB-ELSA~\cite{Crede2005}, LNS-GeV-$\gamma$~\cite{Nakabayashi2006}
and GRAAL~\cite{Bartalini2007}, polarized beam asymmetries by
CB-ELSA/TAPS~\cite{Elsner2007} and GRAAL~\cite{Bartalini2007}.

The copious set of data has motivated extensive theoretical
investigations. Most of the available models are based on
meson-nucleon degrees of freedom, in which the Feynman diagrammatic
techniques are used, so that the transition amplitudes are Lorentz
invariant. In recent years various advanced approaches have been
developed, namely, the unitary isobar model of
MAID~\cite{Chiang2003a}, Geissen~\cite{Geissen} and Bonn-Gatchina
groups~\cite{Sarantsev2005} coupled-channel approaches, as well as
the partial wave analysis of SAID~\cite{Arndt2007}. Those approaches
have no explicit connection with QCD, and the number of parameters
in the models increases with the number of resonances included in
the models.

Formalisms embodying the subnucleonic degrees of freedom are also
being developed. Such a program has its genesis in the early works
by Copley, Karl and Obryk~\cite{Copley1969} and Feynman, Kisslinger
and Ravndal~\cite{Feynman1971} in the pion photoproduction, who
provided the first clear evidence of the underlying $SU(6)\otimes
O(3)$ structure of the baryon spectrum. The subsequent
works~\cite{Koniuk1980,Capstick1992} in the framework of the
constituent quark models concentrated mainly on the transition
amplitudes and the baryon mass spectrum, predicting still
undiscovered or "missing", resonances. However, those approaches did
not investigate reaction mechanisms.

In Ref.~\cite{Li1997} a comprehensive and unified approach to the
pseudoscalar mesons photoproduction, based on the low energy QCD
Lagrangian~\cite{Manohar1984}, is developed with the explicit quark
degrees of freedom. This approach reduces drastically the number of
free parameters, for example, within the exact $SU(6)\otimes O(3)$
symmetry, the reaction under investigation has only one free
parameter, namely, $\eta NN$ coupling constant. However, that
symmetry is broken and in order to take into account that effect,
one free parameter per resonance was introduced in previous
calculations~\cite{Li1998,Saghai2001}. Given that the configuration
mixing among the 3-constituent quarks bound states is a consequence
of the $SU(6)\otimes O(3)$ symmetry breakdown, in the present work
we use the one-gluon-exchange mechanism to generate the
configuration mixing of the wave functions. In this approach, the
number of parameters decreases significantly. After the parameters
are determined by fitting the data, we then study the contributions
from the missing resonances (see e.g.
Refs.~\cite{Capstick2000,Bijker,Gian}). Besides, we give relations
connecting the scattering amplitudes in our $\chi$QM approach to the
photoexcitation helicity amplitudes and partial decay widths of
resonances. Our approach offers also the opportunity of
investigating new nucleon resonances, for which strong indications
have been reported in the
literature~\cite{Sarantsev2005,Saghai2001,Li1996,Batinic1997,Giannini2001,BES,Chen2003,
Chiang2003,Tryasuchev2004,Mart04,Kelkar,Julia2006}.

The paper is organized as follows.  In Section~\ref{Sec:Theo}, the
theoretical content of our work is presented. Starting from a chiral
effective Lagrangian, the CGLN amplitudes for the process
$\gamma~p~\to~\eta~p$ are given within the $SU(6)\otimes O(3)$
symmetry. Then the consequences of the breaking of that symmetry
{\it via} configuration mixing in One-Gluon-Exchange (OGE) model is
reported and helicity amplitudes of photon transition and meson
decay partial widths of resonances are presented. The fitting
procedure and numerical results for differential cross-section,
polarized beam asymmetry, helicity amplitudes, and $N^*~\to~\eta~N$
partial decay width are reported and discussed in
Section~\ref{Sec:Res}, where possible roles played by new / missing
resonances are examined. Summary and conclusions are given in
Section~\ref{Sec:Conclu}.

%
\section{Theoretical Frame}\label{Sec:Theo}

In this Section we recall the content of a chiral constituent quark
approach and relate it to the configuration mixing of constituent
quarks states {\it via} a OGE model, generated by the breakdown of
the $SU(6)\otimes O(3)$ symmetry. Then we present issues related to
the photoexcitation helicity amplitudes and the partial decay widths
of nucleon resonances.

%

\subsection{Chiral constituent quark model}
\label{Sec:QM}

As in Ref.~\cite{Li1997} we start from an effective chiral
Lagrangian~\cite{Manohar1984},
\begin{eqnarray} \label{lg}
\mathcal{L}=\bar{\psi}[\gamma_{\mu}(i\partial^{\mu}+V^{\mu}+\gamma_5A^{\mu})-m]\psi
+\cdot\cdot\cdot,
\end{eqnarray}
\noindent where vector ($V^{\mu}$) and axial ($A^{\mu}$) currents
read,
\begin{eqnarray}
V^{\mu} =
 \frac{1}{2}(\xi\partial^{\mu}\xi^{\dag}+\xi^{\dag}\partial^{\mu}\xi)~,~A^{\mu}=
 \frac{1}{2i}(\xi\partial^{\mu}\xi^{\dag}-\xi^{\dag}\partial^{\mu}\xi),
\end{eqnarray}
with $ \xi=\exp{(i \phi_m/f_m)}$ and $f_m$ the meson decay constant.
$\psi$ and $\phi_m$ are the quark and meson fields, respectively.

There are four components for the photoproduction of pseudoscalar
mesons based on the QCD Lagrangian,
\begin{eqnarray}\label{eq:Mfi}
{\cal M}_{fi}&=&\langle N_f| H_{m,e}|N_i \rangle + \sum_j\bigg \{
\frac {\langle N_f|H_m |N_j\rangle \langle N_j |H_{e}|N_i\rangle
}{E_i+\omega-E_j}+ \nn \\ && \frac {\langle N_f|H_{e}|N_j\rangle
\langle N_j|H_m |N_i\rangle }{E_i-\omega_m-E_j}\bigg \}+{\cal M}_T ,
\end{eqnarray}
where $N_i(N_f)$ is the initial (final) state of the nucleon, and
$\omega (\omega_{m})$ represents the energy of incoming (outgoing)
photons (mesons). The first term in Eq.~(\ref{eq:Mfi}) is a seagull
term. It is generated by the gauge transformation of the axial
vector $A_{\mu}$ in the QCD Lagrangian. This term, being
proportional to the electric charge of the outgoing mesons, does not
contribute to the production of the $\eta$-meson. The second and
third terms correspond to the {\it s-} and {\it u-}channels,
respectively. The last term is the {\it t-}channel contribution.

In this paper we focus on the nucleon resonance contributions. Given
that the {\it u-}channel contributions are less sensitive to the
details of resonances structure than those in the {\it s-}-channel,
it is then reasonable to treat the {\it u-}channel components as
degenerate~\cite{Saghai2001}.

For {\it s-}channel, the amplitudes are given by the following
expression~\cite{Li1997,Saghai2001}:
\begin{equation}
\label{42} {\cal M}_{N^*}=\frac
{2M_{N^*}}{s-M_{N^*}^2-iM_{N^*}\Gamma({\bf q})}e^{-\frac {{\bf
k}^2+{\bf q}^2}{6\alpha^2}} {\cal O}_{N^*},
\end{equation}
where $\sqrt {s}~\equiv~W~=~E_N+\omega_{\gamma}=E_S+\omega_m$ is the
total centre-of-mass energy of the system, and ${\cal O}_{N^*}$ is
determined by the structure of each resonance. $\Gamma({\bf q})$ in
Eq. (\ref{42}) is the total width of the resonance, and a function
of the final state momentum ${\bf q}$.

The transition amplitude for the $n^{th}$ harmonic-oscillator shell
is
\begin{equation}
\label{61} {\cal O}_{n}={\cal O}_n^2 +{\cal O}_n^3.
\end{equation}

The first (second) term represents the process in which the incoming
photon and outgoing meson, are absorbed and emitted by the
same(different) quark.

In the present work, we use the standard multipole expansion of the
CGLN amplitudes~\cite{Chew1957}, and obtain the partial wave
amplitudes of resonance $l_{2I, 2l\pm1}$. Then, the transition
amplitude takes the following form:
\begin{eqnarray}
\label{63} {\cal O}_{N^*}=if_{1l\pm}  {\bf \sigma} \cdot {\bf
\epsilon}+ f_{2l\pm} {\bf \sigma} \cdot {\bf \hat{q}} {\bf \sigma}
\cdot ({\bf \hat{k}} \times {\bf \epsilon})+ if_{3l \pm} {\bf
\sigma} \cdot {\bf \hat{k}} {\bf \hat{q}} \cdot {\bf \epsilon} +
if_{4l\pm}  {\bf \sigma} \cdot {\bf \hat{q}}{\bf \epsilon}\cdot {\bf
\hat{q}}.
\end{eqnarray}
Expressing the CGLN amplitudes in their usual
formulation~\cite{Walker1969,Chaloupka1974}, leads to the
Hebb-Walker amplitudes in terms of photoexcitation helicity
amplitudes,
\begin{eqnarray}
A_{l\pm}&=&\mp   fA^{N^*}_{1/2},\\
B_{l\pm}&=&\pm   f \sqrt{\frac{4}{l(l+2)}}A^{N^*}_{3/2},
\end{eqnarray}
where
\begin{eqnarray}\label{Eq: f}
f=\frac{1}{(2J+1)2\pi}[\frac{M_NE_N}{M_{N^*}^2}k]^{1/2}\frac{2M_{N^*}}{s-M_{N^*}^2+iM_{N^*}\Gamma({\bm
q})}A^m_{1/2}\equiv
f_0\frac{2M_{N^*}}{s-M_{N^*}^2+iM_{N^*}\Gamma({\bm q})},
\end{eqnarray}
with $A^m_{1/2}$ the $N^*~\to~\eta N$ decay amplitude, appearing in
the partial decay width,
\begin{eqnarray}
 \Gamma_m=\frac{1}{(2J+1) }\frac{|{\bf q}|E_N}{\pi
M_{N^*}}|A^m_{1/2}/C^I_{m N}|^2,
\end{eqnarray}
where $C^I_{\pi N}$ represents the Clebsch-Gordan coefficients
related to the isospin coupling in the outgoing channel.

In Ref.~\cite{Li1997}, the partial decay amplitudes are used to
separate the contribution of the state with the same orbital angular
momentum $L$. In fact, with the helicity amplitudes of photon
transition and meson decay we can directly obtain the CGLN
amplitudes for each resonances in terms of Legendre polynomials
derivatives:
\begin{eqnarray}\label{Eq: AA to f}
f_{1l\pm}&=&f_0[ \mp A^{N^*}_{1/2}-\sqrt{\frac{l+ 1/2\mp 1/2}{l+ 1/2
\pm 3/2}}
A^{N^*}_{3/2}]P'_{\ell\pm1},\nonumber\\
f_{2l\pm}&=&f_0[ \mp A^{N^*}_{1/2}-\sqrt{\frac{l+ 1/2\pm 3/2}{l+ 1/2
\mp\ 1/2}}
A^{N^*}_{3/2}]P'_{\ell},\nonumber\\
f_{3l\pm}&=&\pm f_0\frac{2A^{N^*}_{3/2}}{\sqrt{(l- 1/2 \pm 1/2)(l+
3/2 \pm 1/2)}}
P''_{\ell\pm1},\nonumber\\
f_{4l\pm}&=&\mp f_0\frac{2A^{N^*}_{3/2}}{\sqrt{(l- 1/2 \pm 1/2)(l+
3/2 \pm\ 1/2)}}P''_{\ell}.
\end{eqnarray}
All $f_i$s are proportional to the meson decay amplitudes. So they
can be used to separate the contributions from the state with the
same $N$ and $L$ as presented in Ref.~\cite{Li1997}.

In our approach, the photoexcitation helicity amplitudes
$A^{N^*}_{1/2}$ and $A^{N^*}_{3/2}$, as well as the decay
amplitudes, are related to the matrix elements of the
electromagnetic interaction Hamiltonian~\cite{Copley1969},
\begin{eqnarray}A_\lambda  &=&\sqrt{\frac{2\pi}{k}}\langle
{N^*};J\lambda|H_{e}|N;\frac{1}{2}\lambda-1\rangle,  \\
A^m_\nu&=&\langle N;\frac{1}{2}\nu|H_{m}|{N^*};J\nu\rangle.
\end{eqnarray}

%

\subsection{Configuration Mixing}
\label{Sec:mixing}
The amplitudes in Sec.\ref{Sec:QM} are derived under the
$SU(6)\otimes O(3)$ symmetry. However, for physical states that
symmetry is broken. An example is the violation of the Moorhouse
rule~\cite{Moor}. In Ref.~\cite{Li1998}, a set of parameters
$C_{{N^*}}$ were hence introduced to take into account the breaking
of that symmetry, {\it via} following substitution:
\begin{eqnarray}\label{eq:AR}
{\mathcal O}_{N^*} \to C_{N^*} {\mathcal O}_{N^*} .
\end{eqnarray}
In Refs.~\cite{Li1998,Saghai2001}, those parameters were allowed to
vary around their $SU(6)\otimes O(3)$ values ($\vert C_{N^*} \vert$
= 0 or 1). In this work, instead of using those adjustable
parameters, we introduce the breakdown of that symmetry through the
configuration mixings of baryons wave functions.

To achieve such an improvement, we must choose a potential model.
The popular used ones are one-gluon-exchange (OGE)
model~\cite{Isgur1978,Isgur1978a,Isgur1979} and Goldstone boson
exchange model~\cite{Glozman1996}. As shown in
Refs.~\cite{He2003a,Wang2003}, these two models give similar mixing
angles for the negative parity resonances and the relevant
observables. Here, we adopt the OGE model which has been
successfully used to study the helicity amplitudes and decay
widths~\cite{Koniuk1980} of resonances.

In OGE model, the Hamiltonian of system can be written
as~\cite{Isgur1978,Isgur1978a,Isgur1979},
\begin{equation} H=\sum_{i=1}^{3} m_i + \sum_{i=1}^{3}
 \frac{p_i^2}{2m_i^2} +\sum_{i<j=1}^{3}\half K r_{ij}^2+
 \sum_{i<j=1}^{3}U(r_{ij}) +H_{hyp}\
,\label{Eq: Hamiltonian}\end{equation}
where the $m_i$ is the "constituent" effective masse of quark $i$
and $\bm{r_{ij}~=~r_{i}-r_{j}}$ the separation between two quarks.
The confinement potential has two components; one written as a
harmonic oscillator potential ($\half K r_{ij}^2$, with $K$ the
spring constant), and an unspecified anharmonicity $U(r_{ij})$,
treated as a perturbation.

The hyperfine part interaction is the sum of contact and tensor
terms,
\begin{equation}H_{hyp} =\frac{2\alpha_s}{3m_q^2}
\sum_{i<j=1}^{3} \left\{ \ \frac{8\pi}{3}\
 {\bf S}_i
\cdot{\bf S}_j\ \ \delta^3({\bf r}_{ij})+\frac{1}{ {\bm r}_{ij}^3} (
\frac{3{\bf S}_i\cdot{\bf r}_{ij}\ {\bf S}_j\cdot{\bm r}_{ij}}{{\bf
r}_{ij}^2}-{\bf S}_i \cdot{\bf S}_j)\ \right\}.
\end{equation}
Here, ${\bf S}_i$ is the spin of quark $i$, and $\alpha_s$ a
normalization factor, treated as free parameter~\cite{Isgur1978a}.

The hyperfine interaction generates the configuration mixings among
the ground-state $N^2S_S$ ($[56, O^+]$) and other configurations,
e.g. $N^2S'_S$ ($[56', O^+]$), $N^2S_M$ ($[70, O^+]$), and $N^4D_M$
($[70, 2^+]$). Here, the notation is $X^{2S+1}L_\pi$, where
$X~=~N,~\Delta,~\Sigma,...$, $S$ the total quark spin, $L$ $=$ $S$,
$P$, $D...$ the total orbital angular momentum, and $\pi~=~S$,~$M$
or $A$ the permutational symmetry (symmetric, mixed symmetry, or
antisymmetric, respectively) of spatial wave function.

The first two terms in Eq.~(\ref{Eq: Hamiltonian}) can be rewritten
as two harmonic oscillators within the Jocabi coordinate. Its
solution is the well known $SU(6)\otimes O(3)$ wave functions. The
breakdown of the symmetry arises from the additional terms. Given
that the configuration mixing is mainly produced by the spin- and
flavor-dependent parts of Hamiltonian~\cite{He2003a}, here we use a
simple method to deal with the confinement terms in
Refs.~\cite{Isgur1979,Capstick2000}, where three constants $E_0$,
$\Omega$, and $\Delta$ are introduced.

In order to illustrate the modifications of the scattering
amplitudes due to the $SU(6)\otimes O(3)$ symmetry breakdown, we
give in the following the explicit derivations in the case of the
$S_{11}(1535)$ resonance . In lines with Ref.~\cite{Saghai2001}, we
express the amplitudes ${\mathcal A}_{S_{11}}$ in terms of the
product of the photoexcitation and meson-decay transition
amplitudes,
\begin{eqnarray}
\label{eq:MixAR1} {\mathcal A}_{S_{11}} \propto <N|H_m|
S_{11}><S_{11}|H_e|N>,
\end{eqnarray}
where $H_m$ and $H_e$ are the meson and photon transition operators,
respectively.  The wave function can be written within the
$SU(6)\otimes O(3)$ symmetry for $n \le 2$ shells as
$X^{2S+1}L_\pi{J^P}$ and configuration mixings, with $J^P$ the
state's total angular momentum and parity,
\begin{eqnarray}
|S_{11}(1535)\rangle&=&-\sin\theta_{S}|N^4P_M{\textstyle{1\over
2}^-}\rangle+
\cos\theta_{S}|N^2P_M{\textstyle{1\over 2}^-}\rangle,\\
|Nucleon\rangle &=& c_1|N^2S_S{\textstyle{1\over 2}^+}\rangle+
c_2|N^2S'_S{\textstyle{1\over
2}^+}\rangle+c_3|N^4D_M{\textstyle{1\over 2}^+}\rangle+ \nn \\ &&
c_4|N^2S_M{\textstyle{1\over
2}^+}\rangle+c_5|N^2P_A{\textstyle{1\over 2}^+}\rangle,\end{eqnarray}
where $\theta_{S}$ and $c_i$ can be determined by the OGE model. If
we set $c_1=1$ and $c_{2,3,4,5}=0 $ (so $\theta_S=0$), then, the
$SU(6)\otimes O(3)$ symmetry is restored. The improvement compared
to Ref.~\cite{Saghai2001} is that here we not only take into account
the mixing in the intermediate $S_{11}$ resonance but also in the
initial- and final-state nucleon. Moreover, for other resonances, we
also include directly the configuration mixing of wave functions
{\it via} OGE model, so that we do not need to introduce the free
parameters $C_{{N^*}}$ (Eq.~(\ref{eq:AR})).

The electromagnetic transition amplitudes then take the following
form:
\begin{eqnarray}\label{eq:ci}
<S_{11}|H_e|N>& =&c_1<S_{11}|H_e|N^2S_S{\textstyle{1\over
2}^+}\rangle+ c_2<S_{11}|H_e|N^2S'_S{\textstyle{1\over
2}^+}\rangle+c_3<S_{11}|H_e|N^4D_M{\textstyle{1\over
2}^+}\rangle\nonumber\\&+&c_4<S_{11}|H_e|N^2S_M{\textstyle{1\over
2}^+}\rangle+c_5<S_{11}|H_e|N^2P_A{\textstyle{1\over 2}^+}\rangle\nonumber\\
&=&c_1\cos\theta<N^2P_M{\textstyle{1\over
2}^-}|H_e|N^2S_S{\textstyle{1\over 2}^+}\rangle+...
\end{eqnarray}
Here, the term $<N^4P_M{\textstyle{1\over
2}^-}|H_e|N^2S_S{\textstyle{1\over 2}^+}\rangle$ vanishes because of
the Moorhouse rule~\cite{Moor}. In Ref.~\cite{Saghai2001}, the
mixing angles are introduced also to give a nonzero value for
contributions from the $D_{13}(1700)$ resonance, but the nucleon
wave function includes only the $n=0$ part, that is, $c_1=1$,
$c_{2,3,4,5}=0$. Moreover, the contribution of the $D_{15}(1675)$
($|N^4D_M{\textstyle{5\over 2}^+}\rangle$ state) is zero, if we
consider only the wave function up to $n=2$. Then, in
Ref.~\cite{Saghai2001}, for this latter resonance a term identical
to the contribution to the $\eta$ photoproduction on neutron target
was added by hands. In this work, the nucleon wave function with
$n=2$ produces {\it naturally} a non-zero contribution with the same
form as for neutron target under the $SU(6)\otimes O(3)$ symmetry.

Analogously, for meson decay amplitudes we get,
\begin{eqnarray}
<N|H_m| S_{11}>& =&c_1(\cos\theta _{S} - {\cal {R}} \sin \theta
_{S})<N^2S_S{\textstyle{1\over 2}^+}|H_m|N^2P_M{\textstyle{1\over
2}^-}\rangle+...
\end{eqnarray}
and the ratio
\begin{eqnarray}\label{eq:MixR}
{\cal {R}} =  \frac {<N|H_m|N(^4P_M)_{{\frac 12}^-}>}
{<N|H_m|N(^2P_M)_{{\frac 12}^-}>},
\end{eqnarray}
is a constant determined by the $SU(6)\otimes O(3)$ symmetry.

Then, Eq.~(\ref{eq:MixAR1}) reads,
\begin{eqnarray}
\label{eq:MixAR2} {\mathcal A}_{S_{11}}=C_{S_{11}}
<N^2S_S{\textstyle{1\over 2}^+}|H_m| N^2P_M{\textstyle{1\over
2}^-}><N^2P_M{\textstyle{1\over 2}^-}|H_e|N^2S_S{\textstyle{1\over
2}^+}>+...,
\end{eqnarray}
where,
\begin{eqnarray}C_{S_{11}}=c_1^2(\cos^2 \theta _{S} - {\cal {R}} \sin
\theta _{S}\cos \theta _{S})+...
\end{eqnarray}

So, if we remove all $n=2$ parts from the wave function of the
nucleon, as in Ref.~\cite{Li1998}, then the factor $C_{S_{11}}$ is a
constant. However after other contributions are included, it becomes
dependent on the momenta $\bm k$ and $\bm q$. In this work we keep
this dependence.

%

\section{Results and discussion}\label{Sec:Res}

With the formalism presented in Sec.\ref{Sec:Theo}, we investigate
the process $\gamma p \to \eta p$. A chiral constituent quark model
was proven~\cite{Saghai2001} to be an appropriate approach to that
end. That work embodied one free parameter per nucleon resonance, in
order to take into account the breaking of the $SU(6) \otimes O(3)$
symmetry. In the present work, this latter phenomenon is treated via
configuration mixing, reducing the number of adjustable parameters.
As in Refs.~\cite{Saghai2001}, we introduce resonances in $n \le 2$
shells, to study the $\eta$ photoproduction in the centre-of-mass
energy $W~\le 2$ GeV.

%
\subsection{Fitting procedure}\label{Sec:Fit}
Using the CERN MINUIT code, we have fitted simultaneously the
following data sets:

\begin{itemize}

 \item {\bf Differential cross-section:} Data base includes 1220 data points,
for 1.49 $\lesssim~W~\le$ 1.99 GeV, coming from the following labs:
MAMI~\cite{Krusche1995}, CLAS~\cite{Dugger2002},
ELSA~\cite{Crede2005}, LNS~\cite{Nakabayashi2006}, and
GRAAL~\cite{Bartalini2007}. Only statistical uncertainties are used.

 \item {\bf Polarized beam asymmetry:} Polarized beam asymmetries (184 data points),
for 1.49 $\lesssim~W~\le$ 1.92 GeV,
 from GRAAL~\cite{Bartalini2007} and ELSA~\cite{Elsner2007}. Only
statistical uncertainties are used.
 \item {\bf Spectrum of known resonances:} For spectrum of known resonances,
 we use as input their PDG values~\cite{Yao2006} for masses and widths, with the
uncertainties reported there plus an additional theoretical
uncertainty of 15 MeV, as in Ref.~\cite{Capstick1992}, in order to
avoid overemphasis of the resonances with small errors. The data
base contains
 all 12 known nucleon resonances
 as in PDG, with $M~\le$~2 GeV, namely,

{\boldmath$ n$}{\bf =1:} $S_{11}(1535)$, $S_{11}(1650)$,
$D_{13}(1520)$, $D_{13}(1700)$, and $D_{15}(1675)$;

{\boldmath$ n$}{\bf =2:} $P_{11}(1440)$, $P_{11}(1710)$,
 $P_{13}(1720)$, $P_{13}(1900)$,
$F_{15}(1680)$, $F_{15}(2000)$, and $F_{17}(1990)$.

Besides the above isospin-1/2 resonances, we fitted also the mass of
$\Delta$(1232) resonance. However, spin-3/2 resonances do not
intervene in the $\eta$ photoproduction.
 \item {\bf Additional resonance:} Resonances with masses above $M~\approx$~2 GeV,
treated as degenerate, are simulated by a single resonance, for
which are left as adjustable parameters the mass, the width, and the
symmetry breaking coefficient.
\end{itemize}

The adjustable parameters, listed in Table~\ref{Tab:GBE_1}, are as
follows: $\eta$ nucleon coupling ($g_{\eta NN}$), mass of the
non-strange quarks ($m_q$), harmonic oscillator strength ($\alpha$),
QCD coupling constant ($\alpha_s$), confinement constants ($E_0$,
$\Omega$, and $\Delta$), three parameters $M$, $\Gamma$, and
$C_{N}^*$ related to the degenerate treatment of resonances with
masses above $\approx$ 2 GeV, and the strength of the $P_{13}(1720)$
resonance. We will come back to this latter parameter.

The spectrum of the known resonances put constraints on six of the
adjustable parameters. Five of them ($m_q$, $\alpha$, $\alpha_s$,
$\Omega$, and $\Delta$) are determined through an interplay between
the mass spectrum of the resonances and the photoproduction data
{\it via} the configurations mixings parameters $c_i$
(Eq.~\ref{eq:ci}). The sixth one, $E_0$, is determined by the mass
of nucleon. The coupling constant $g_{\eta NN}$ is determined by
photoproduction data. The parameter $C_{P_{13}(1720)}$ is the
strength of the ${P_{13}(1720)}$ resonance, that we had to leave as
a free parameter in order to avoid its too large contribution
resulting from direct calculation. This latter parameter, as well as
those defining the higher mass resonance (HM $N^*$) are determined
by the photoproduction data. Notice that in fitting the
photoproduction data, we use the PDG~\cite{Yao2006} values for
masses and widths of resonances.

The complete set of adjustable parameters mentioned above, leads to
our model $A$ (see $3^{rd}$ column in Table~\ref{Tab:GBE_1} for
which the reduced $\chi^2$ turns out to be large (12.37).

In recent years, several
authors~\cite{Sarantsev2005,Saghai2001,Li1996,Batinic1997,Giannini2001,BES,Chen2003,
Chiang2003,Tryasuchev2004,Mart04,Kelkar,Julia2006} have put forward
need for new resonances in interpreting various observables, with
extracted masses roughly between 1.73 and 2.1 GeV. We have hence,
investigated possible contributions from three of them: $S_{11}$,
$D_{13}$, and $D_{15}$. For each of those new resonances we
introduce then three additional adjustable parameters per resonance:
mass~($M$), width~($\Gamma$), and symmetry breaking
coefficient~($C_{N^*}$). Fitting the same data base, we obtained a
second model, called model $B$, for which the adjustable parameters
are reported in the last column of Table~\ref{Tab:GBE_1}. The
reduced $\chi^2$ is very significantly improved going down from
12.37 to 2.31. In the rest of this Section, we concentrate on the
model $B$.

\begin{table*}[ht!]

\caption{Adjustable parameters and their extracted values, with
$m_q$, $\alpha$, $E_0$, $\Omega$, $\Delta$, $M$, and $\Gamma$ in
MeV.}
\renewcommand\tabcolsep{0.2cm}
\begin{tabular}{llrr}  \hline\hline

  & Parameter            & Model $A$ & Model $B$ \\ \hline

  & $g_{\eta NN}$      & 0.391 & 0.449 \\
  & $m_q$              & 277   & 304   \\
  & $\alpha$           & 288   & 285   \\
  & $\alpha_s$         & 1.581 & 1.977 \\
  & $E_0$              & 1135  & 1138  \\
  & $\Omega$           & 450   & 442   \\
  & $\Delta$           & 460   & 460   \\
  & $C_{P_{13}(1720)}$ & 0.382 & 0.399 \\
HM $N^*$: &                       & & \\
              &    $M$          & 1979 & 2129 \\
              &    $\Gamma$     &  124 &   80 \\
              &    $C_{N^*}$    & -0.85 & -0.70 \\
New $S_{11}$: &                       & & \\
              &    $M$                & & 1717 \\
              &    $\Gamma$           & &  217 \\
              &    $C_{N^*}$          & & 0.59 \\
New $D_{13}$: &                       & &       \\
              &    $M$                & & 1943 \\
              &    $\Gamma$           & &  139 \\
              &    $C_{N^*}$          & & -0.19 \\
New $D_{15}$: &                       & &       \\
              &    $M$                & & 2090 \\
              &    $\Gamma$           & &  328 \\
              &    $C_{N^*}$          & & 2.89  \\
\hline
$\chi^2_{d.o.f}$ &                       & 12.37 & 2.31 \\
\hline \hline\end{tabular} \label{Tab:GBE_1}

\end{table*}

Extracted values within OGE model come out close to those used by
Isgur and Capstick~\cite{Isgur1979,Capstick2000}: $E_0$~=~1150 MeV,
$\Omega$ $\approx$ 440 MeV, $\Delta$ $\approx$ 440 MeV. For three
other parameters, Isgur and
 Capstick introduce $\delta$ = $(4 \alpha_s \alpha) / (3 \sqrt{2 \pi} m_{u}^2)$,
 for which they get $\approx$ 300 MeV. Model $B$ gives $\delta$ $\approx$ 262 MeV.

For the three new resonances, we follow the method in
Ref.~\cite{Li1998}, as discussed in Sec.\ref{Sec:mixing}, {\it via}
Eq.~(\ref{eq:AR}). The extracted Wigner mass and width, as well as
the strength for those resonances are given in
Table~\ref{Tab:GBE_1}.

For the new $S_{11}$, the Wigner mass and width are consistent with
the values in Refs.~\cite{Saghai2001,Li1996,Batinic1997,Julia2006},
but the mass is lower, by about 100 to 200 MeV, than findings by
other
authors~\cite{Capstick1994,Giannini2001,Chen2003,Chiang2003,Tryasuchev2004}.
The most natural explanation would be that it is the first $S_{11}$
state in the $n=3$ shell, however its low mass could indicate a
multiquark component, such as, a quasi-bound
kaon-hyperon~\cite{Li1996} or a pentaquark state~\cite{Zou2007}. For
the $D_{13}(1850)$, the variation of $\chi^2$ is small.
Interestingly, we find large effect from a $D_{15}$ state around
2090 GeV with a Wigner width of 330 MeV. It is very similar to the
$N(2070)D_{15}$ reported in Refs.~\cite{Crede2005,Sarantsev2005}. It
can be explained as the first $D_{15}$ state in $n=3$
shell~\cite{Crede2005}.

The results of baryon spectrum extracted from the present work are
reported in Tables~\ref{Tab:MS1} and~\ref{Tab:MS2}.
Table~\ref{Tab:MS1} is devoted to the known resonances. Our results
are in good agreement with those obtained by Isgur and
Karl~\cite{Isgur1978a,Isgur1979}, and except for the $S_{11}$(1535)
and $D_{13}$(1520), fall in the ranges estimated by
PDG~\cite{Yao2006}. The additional "missing" resonances generated by
the OGE model, are shown in Table~\ref{Tab:MS2}. The extracted
masses are compatible with those reported by Isgur and
Karl~\cite{Isgur1978a,Isgur1979}.


\begin{table*}[ht!]

\caption{Extracted masses for known resonances. For each resonance,
results of the present work ($M^{OGE}$) are given in the first line,
predictions from Isgur and Karl for
negative-parity~\cite{Isgur1978a} and
positive-parity~\cite{Isgur1979} excited baryons in the second line,
and PDG values~\cite{Yao2006} in the third line.}
\begin{tabular}{ lcccccc }  \hline\hline

       & $S_{11}$(1535)&$S_{11}$(1650)     & $P_{11}$(1440)   &$P_{11}$(1710)&$P_{13}$(1720)&$P_{13}$(1900)\\ \hline
$M^{OGE}$                        & 1473 & 1620 & 1428 & 1723 & 1718 & 1854 \\
Refs.~\cite{Isgur1978a,Isgur1979}& 1490 & 1655 & 1405 & 1705 & 1710 & 1870 \\
$M^{PDG}$&$1535 \pm 10$&$1655^{+15}_{-10}$&$1440^{+30}_{-20}$&$1710
\pm 30$&$1720^{+30}_{-20}$& $1900$ \\\hline

       & $D_{13}$(1520)&$D_{13}$(1700)&$D_{15}$(1675)&$F_{15}$(1680)&$F_{15}$(2000)&$F_{17}$(1990)\\ \hline
$M^{OGE}$                        & 1511 & 1699 & 1632 & 1723 & 2008 & 1945 \\
Refs.~\cite{Isgur1978a,Isgur1979}& 1535 & 1745 & 1670 & 1715 & 2025 & 1955 \\
$M^{PDG}$&$1520 \pm 5$& $1700 \pm 50$ & $1675\pm 5  $ & $1685 \pm 5 $ & $2000$ & $1990$ \\
\hline \hline
\end{tabular}
\label{Tab:MS1}

\end{table*}


\begin{table*}[ht!]

\caption{Predicted masses for "missing" negative parity excited
baryon by the present work ($M^{OGE}$) and by Isgur and
Karl~\cite{Isgur1979}.}
\renewcommand\tabcolsep{0.16cm}
\begin{tabular}{ lcccccc }  \hline\hline

       & $P_{11}$&$P_{11}$& $P_{13}$&$P_{13}$&$P_{13}$&$F_{15}$\\ \hline
$M^{OGE}$            & 1899 & 2051 & 1942 & 1965 & 2047 & 1943 \\
Ref.~\cite{Isgur1979}& 1890 & 2055 & 1955 & 1980 & 2060 & 1955 \\
\hline \hline
\end{tabular}
\label{Tab:MS2}

\end{table*}


In Table~\ref{Tab:chi2-1}, we examine the sensitivity of our model
to its ingredients by switching off one resonance at a time and
noting the $\chi^2$, without further minimizations. As expected, the
most important role is played by the $S_{11}$(1535), and the effects
of $S_{11}$(1650) and $D_{13}$(1520) turn out to be very
significant. Within the known resonances, the other two ones
contributing largely enough are $F_{15}$(1680) and $P_{13}$(1720).
In addition to those five known resonances, a new $S_{11}$ appears
to be strongly needed by the data, while the smaller effect of a new
$D_{15}$ is found non-negligible. Finally, higher mass resonance ($M
\gtrsim$~2~GeV) and a new $D_{13}$ do not bring in significant
effects.


\begin{table*}[ht!]

\caption{ The $\chi^2$s shown are the values after turning off the
corresponding (known) resonance contribution within the model $B$,
for which $\chi^2$~=~2.31.}
\renewcommand\tabcolsep{0.16cm}
\begin{tabular}{ lcccccc }  \hline\hline

Removed $N^*$& $S_{11}$(1535)&$S_{11}$(1650)&
$P_{11}$(1440)&$P_{11}$(1710)&$P_{13}$(1720)&$P_{13}$(1900)\\ \hline
$\chi^2$     & 162           & 11.9         & 2.29          & 2.39
& 4.15 & 2.35 \\ \hline

Removed $N^*$ &
$D_{13}$(1520)&$D_{13}$(1700)&$D_{15}$(1675)&$F_{15}$(1680)&$F_{15}$(2000)&$F_{17}$(1990)\\
\hline $\chi^2$& 9.83       & 2.29          & 2.24          & 4.82
& 2.33 & 2.31 \\\hline

Removed $N^*$ & HM $N^*$ & New $S_{11}$& New $D_{13}$&New $D_{15}$&
& \\ \hline
$\chi^2$      & 2.50       & 12.69          & 2.63          & 3.88   &  &  \\
\hline \hline
\end{tabular}
\label{Tab:chi2-1}

\end{table*}


Our model $B$ is built upon resonances given in
Table~\ref{Tab:chi2-1}. In Table~\ref{Tab:chi2-2} we investigate
possible contributions from the missing resonances
(Table~\ref{Tab:MS2}). Here, we add them one by one to the model
$B$, without further minimizations. As reported in
Table~\ref{Tab:chi2-2}, none of them play a noticable role in the
reaction mechanism. Please notice that for those resonances we use
the masses that we have determined. We have checked the changes of
the $\chi^2$ by varying those masses by $\pm$100 MeV. Moreover,
given that there is no unique information available on their widths,
we have let them vary between 100 MeV and 1 GeV. The effects of
those procedures on the reported $\chi^2$s in Table~\ref{Tab:chi2-2}
come out to be less than 10\%.
\begin{table*}[ht!]

\caption{The $\chi^2$s shown are the values after adding the
corresponding (missing) resonance contribution within model $B$, for
which $\chi^2$~=~2.31.}
\renewcommand\tabcolsep{0.16cm}
\begin{tabular}{ lcccccc }  \hline\hline

Added $N^*$ & $P_{11}$(1899)&$P_{11}$(2051)&
$P_{13}$(1942)&$P_{13}$(1965)&$P_{13}$(2047)&$F_{15}$(1943)\\ \hline
$\chi^2$    & 2.31 & 2.31 & 2.26 & 2.31 & 2.32 & 2.28 \\
\hline \hline
\end{tabular}
\label{Tab:chi2-2}

\end{table*}


After having discussed above the {\it s-}channel contribution, we
end this Section with a few comments. In our models, non-resonant
components include nucleon pole term, and {\it u-}channel
contributions, treated as degenerate to the harmonic oscillator
shell $n$. {\it t-}channel contributions due to the $\rho$- and
$\omega$-exchanges~\cite{Nimai}, found~\cite{He2007} to be
negligible, are not include in the present work. Our finding about
the effect of higher mass resonances being very small, supports the
neglect of the {\it t-}channel, due to the duality hypothesis (see
e.g. Refs.~\cite{Saghai2001,Duality}).

Finally, the target asymmetry ($T$) data~\cite{Bock1998} are not
included in our data base. Actually, those 50 data points bear too
large uncertainties to put significant constraints on the
parameters~\cite{He2007}.

%
\subsection{Differential cross section and Beam asymmetry}\label{Sec:Diff}

\begin{figure}[ht!]


  \includegraphics[ bb=300 60 550 560 ,scale=0.63 ]{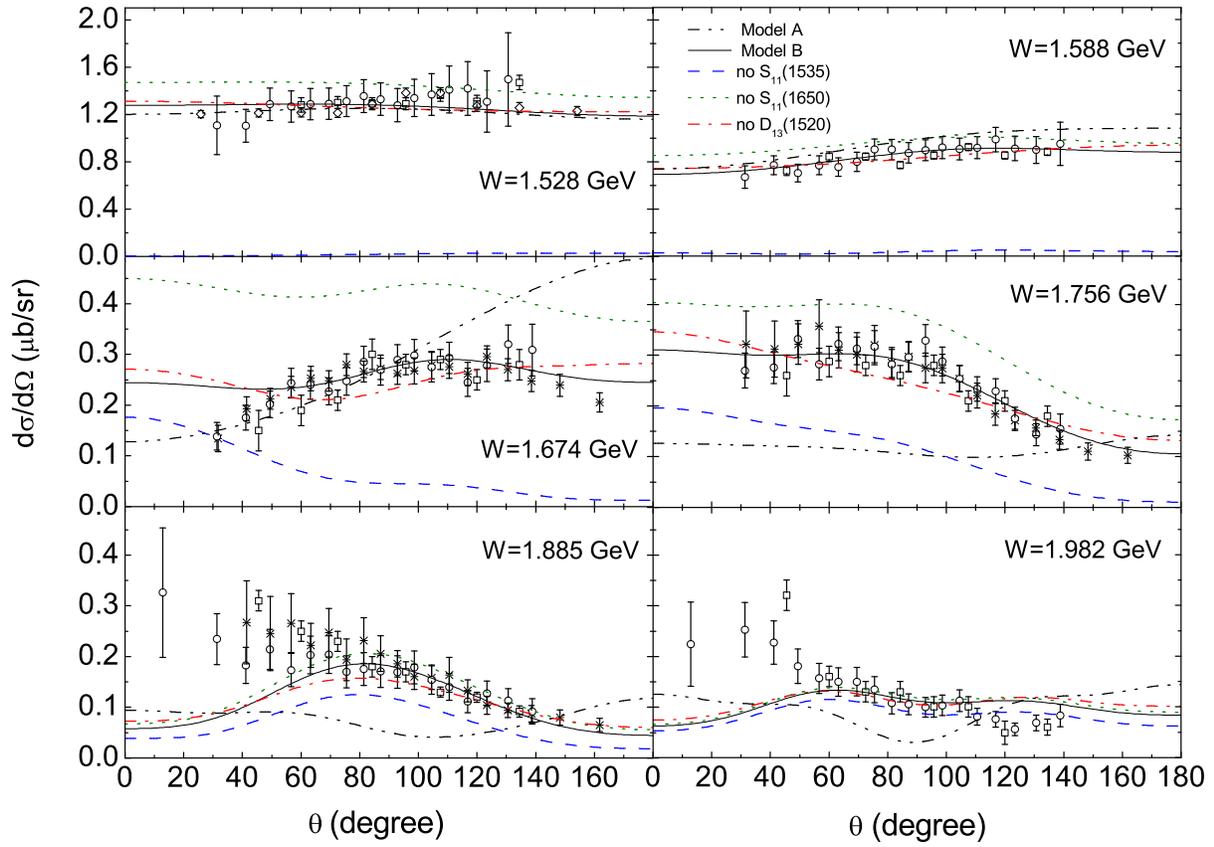}

\caption{Differential cross section for the process
$\gamma~p~\to~\eta~p$. The curves are for models $A$
(dash-dot-dotted) and $B$ (full). The other curves are obtained
within model $B$ by switching off one resonances at a time:
$S_{11}$(1535) (dashed), $S_{11}(1650)$ (dotted) and $D_{13}(1520)$
(dash-dotted). The data are from CLAS (squares)~\cite{Dugger2002},
ELSA (circles)~\cite{Crede2005}, Mainz
(diamonds)~\cite{Krusche1995}, and GRAAL
(stars)~\cite{Bartalini2007}. \label{Fig:ds1}}
\end{figure}

\begin{figure}[ht!]


  \includegraphics[bb=300 60 550 560 ,scale=0.63]{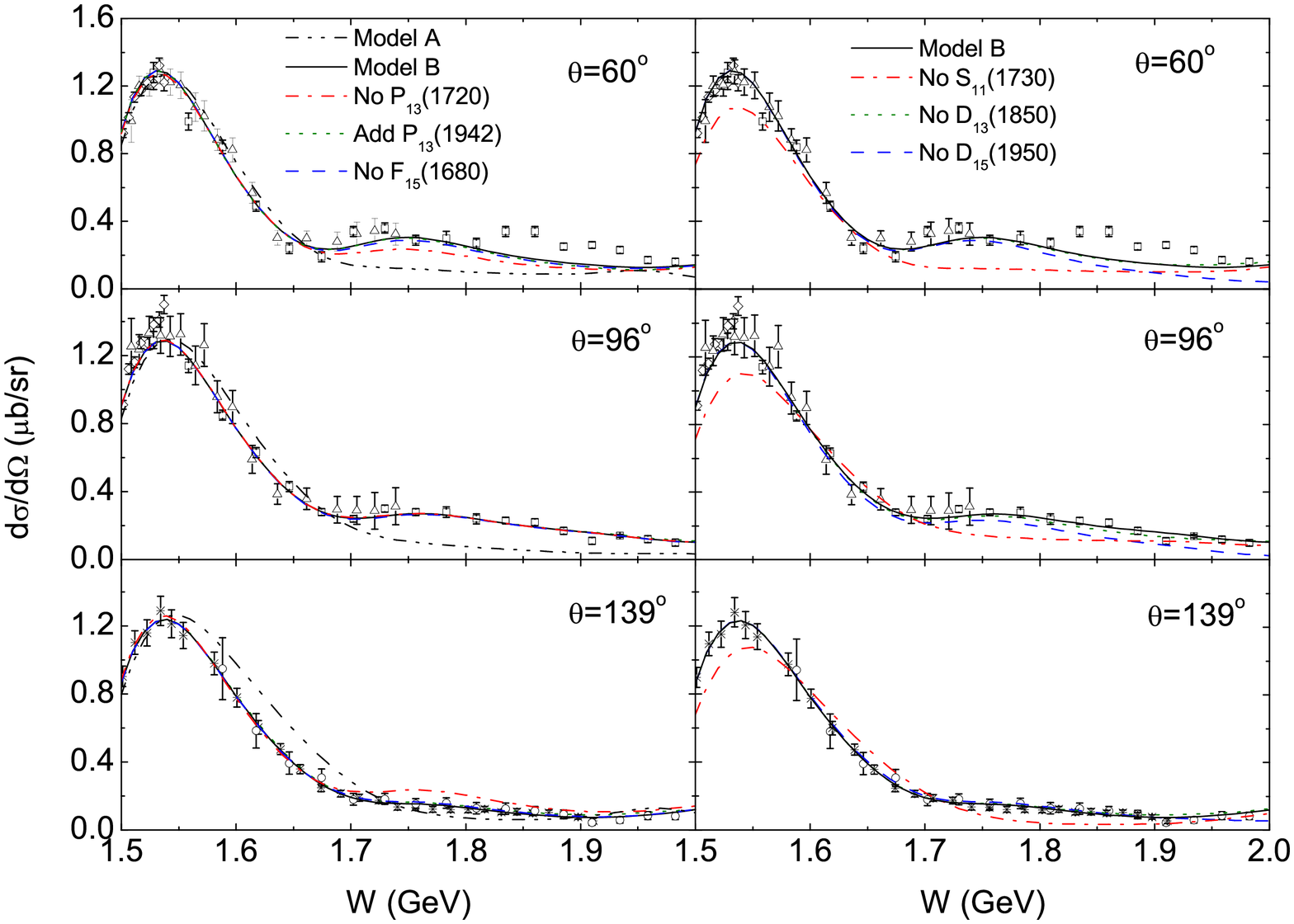}

\caption{Differential cross section as a function of $W$ at three
angles. The dash-dot-dotted and full curves correspond to the models
$A$ and $B$. All other curves are obtained within the model $B$ by
turning off one known resonance or adding a missing one.
 In the left panel
switching off $P_{13}(1720)$ (dash-dotted), $F_{15}(1680)$ (dashed);
adding $P_{13}$(1942) (dotted).  %
In the right panel: switching off $S_{11}(1730)$ (dash-dotted),
$D_{13}(1850)$ (dotted), $D_{15}(1950)$ (dashed). The data are from
 CLAS (squares)~\cite{Dugger2002}, Mainz (diamonds)~\cite{Krusche1995},
LNS (uptriangles)~\cite{Nakabayashi2006}. \label{Fig:ds2}}
\end{figure}

\begin{figure}[ht!]


  \includegraphics[ bb=300 60 550 560 ,scale=0.63  ]{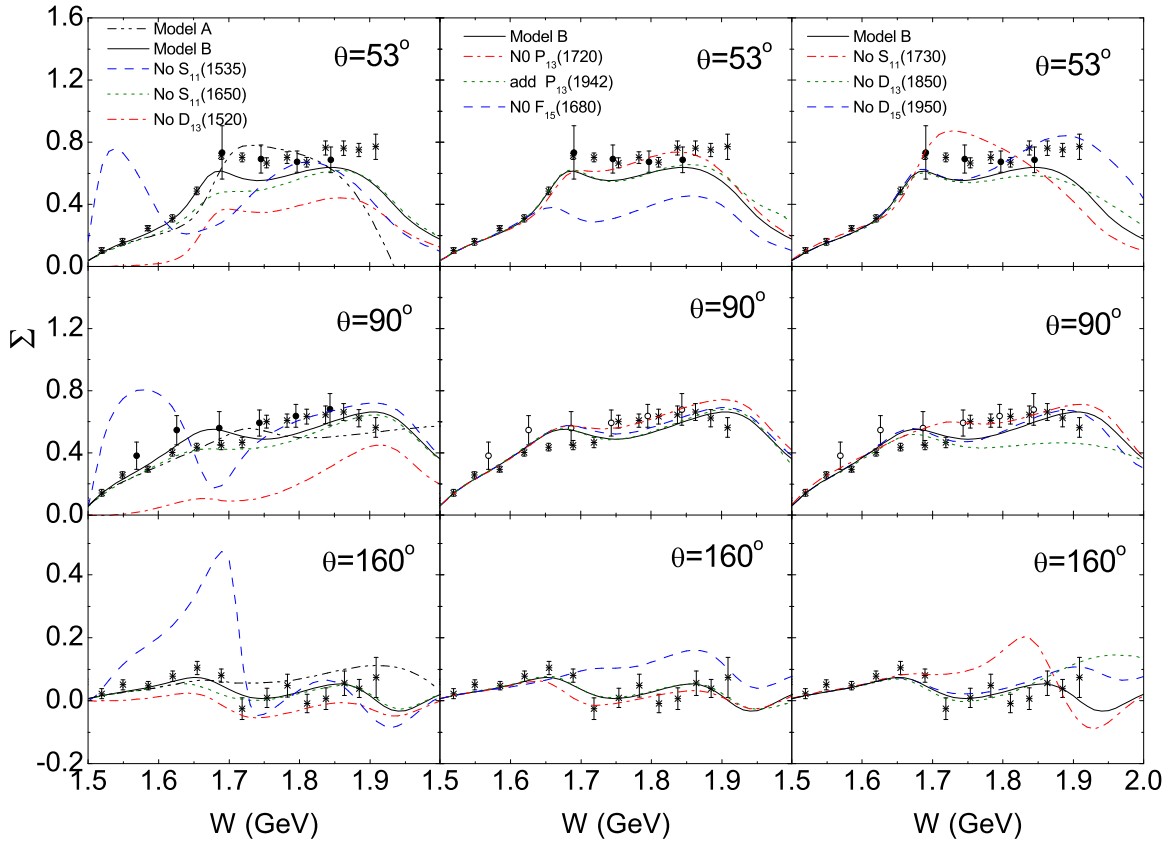}

\caption{Polarized beam asymmetry as a function of $W$. The curves
in the left panel are as in Fig.~\ref{Fig:ds1}, and those in middle
and right pannels as in Fig.~\ref{Fig:ds2}. The data are from ELSA
(full circles)~\cite{Elsner2007} and GRAAL
(stars)~\cite{Bartalini2007}.\label{Fig:pol}}
\end{figure}

In Figures~\ref{Fig:ds1},~\ref{Fig:ds2}, and~\ref{Fig:pol}, we
report our results for angular distributions of differential cross
sections, excitation functions, and polarized beam asymmetries
($\Sigma$), respectively. Results for the models $A$ and $B$ are
shown on all three Figures. The first striking point is that model
$A$ compares satisfactorily with data up to $W~\lesssim$ 1.65 GeV,
but shows very serious shortcomings above, especially in the range W
$\approx$ 1.7 GeV to 1.8 GeV. Model $B$ reproduces the differential
cross section and polarization data well enough, though some
discrepancies appear at the highest energies and most forward angles
($W \gtrsim$  1.85 and $\theta \lesssim$ 50$^\circ$).


In Fig.~\ref{Fig:ds1}, we concentrate on the role played by the
three most relevant known resonances discussed in Sec.\ref{Sec:Fit}
(see Table~\ref{Tab:chi2-1}), namely, by removing one resonance at a
time, within the model $B$. The $S_{11}$(1535) is by far the most
dominant resonance at lower energies and has sizeable effect up to
$W~\approx$ 1.8 GeV, while the $S_{11}$(1650) shows significant
contributions only at intermediate energies. The $D_{13}$(1520) has
less significant contribution, but its role is crucial in
reproducing the correct shape of the differential cross section,
especially at intermediate energies.


The importance of the other two known resonances, leading to a
significant increase of $\chi^2$ when switched off (see
Table~\ref{Tab:chi2-1}), are illustrated in the left panel of
Fig.~\ref{Fig:ds2}. While, the $P_{13}$(1720) affects extreme angles
around $W \approx$ 1.8 GeV, the $F_{15}$(1680) is visible only at
forward angle.

The right panel of Fig.~\ref{Fig:ds2} is devoted to the roles played
by the three new resonances. As mentioned above, the main
shortcoming of the model $A$ appears around $W \approx$ 1.7 - 1.8
GeV. This undesirable feature is cured in the model $B$, due mainly
to the new $S_{11}$, the mass of which turns out to be $M$ = 1.717
GeV. Fig.~\ref{Fig:ds2} illustrates the increase of $\chi^2$
(Table~\ref{Tab:chi2-1}) when that resonance is switched off in the
model $B$. Smaller contributions from the new $D_{15}$ appear in the
forward hemisphere, while the new $D_{13}$ has no significant
manifestation.

Polarized beam asymmetry results are reported in Fig.~\ref{Fig:pol}.
As shown in the left panel of that figure, although the model $B$
gives a better account of the data than the model $A$, the contrast
is less important compared to the differential cross-section
observable. The $S_{11}$(1535) continues playing a primordial role,
while the effect of $S_{11}$(1650) tends to be marginal. This is
also the case (middle panel) for the known $P_{13}$(1720) and
missing $P_{13}$(1942). The established importance of the
$D_{13}$(1520) and $F_{15}$(1680) (in left and middle panels,
respectively) within this observable appear clearly.

In the right panel of Fig.~\ref{Fig:pol}, we examine the case of
three new resonances. The new $S_{11}$ gives sizeable contributions
in the energy range corresponding roughly to its mass. In contrast
to the differential cross-section, the new $D_{13}$ appears to be
significant in the backward hemisphere. Finally, switching off the
new $D_{15}$ improves the agreement with the data at most backward
angles shown, while for the cross-section we get an opposite
behavior. This isolated contradiction reflects the relative weight
of data for the two observables (roughly 6 times more differential
cross-section data than polarization asymmetry, with comparable
accuracies).

This Section, devoted to the observables of the the process
$\gamma~p~\to~\eta~p$, in the energy range $W~~\lesssim$~2 GeV,
leads to the conclusion that within our approach, the reaction
mechanism is dominated by five known and two new nucleon resonances.
%
%
\subsection{Helicity amplitudes and partial decay width}
\label{Sec:Heli}

As discussed in Sec. IV (Eqs. (20), (21), and (25)), our approach
allows calculating the helicity amplitudes and the partial decay
width $N^*~\to~\eta~N$ within a given model without further
adjustable parameters.

\begin{table*}[!ht]

\caption{Helicity amplitudes and decay widths for resonances, with
$\Gamma_{\eta N}^{PDG}=\Gamma_{tot}\cdot Br_{\eta N}$ in PDG
~\cite{Yao2006}.}
\renewcommand\tabcolsep{0.16cm}
\begin{tabular}{ l  c lr rr r }  \hline\hline
Resonances     & $A_{1/2}$   & $A_{1/2}^{PDG}$  &   $A_{3/2}$  &
$A_{3/2}^{PDG}$  & $\sigma\sqrt{\Gamma_{\eta N}}$ &
$(\sigma)\sqrt{\Gamma_{\eta N}^{PDG}}$ \\ \hline
$S_{11}$(1535) & 72 & 90 $\pm$ 30   &   &                           &   7.05&  $(+) 8.87^{+ 1.37}_{-1.37}$   \\
$S_{11}$(1650) & 60 & 53 $\pm$ 16   &   &                           &  -2.20&     $ 1.95^{+ 0.94}_{-1.57}$  \\
$P_{11}$(1440) & 37 &-65 $\pm$ 4    &   &                           &       &                               \\
$P_{11}$(1710) & 27 &   9 $\pm$ 22  &   &                           &   1.30&     $ 2.49^{+ 1.75}_{-0.88}$  \\
$P_{11}$       & 3 &                    &   &                           &  -1.64&                                 \\
$P_{11}$       &-2 &                    &   &                           &  -0.76&                            \\
$P_{13}$(1720) & 194 & 18 $\pm$ 30  &-72&  -19 $\pm$ 20     &   2.07&     $ 2.83^{+ 1.04}_{-0.71}$  \\
$P_{13}$(1900) & 33 &                    &  1&                           &  -0.87&     $ 8.35^{+ 2.11}_{-2.20}$  \\
$P_{13}$       & 32 &                    & -2&                           &   1.80&                                 \\
$P_{13}$       & 14 &                    &  2&                           &   0.05&                              \\
$P_{13}$       & -4 &                    &  4&                           &  -0.73&                                  \\
$D_{13}$(1520) & -20 & -24 $\pm$  9  &144&  166 $\pm\ $5        &   0.30&     $ 0.51^{+ 0.07}_{-0.06}$\\
$D_{13}$(1700) & -6 & -18 $\pm$ 13  &  2&   -2 $\pm$ 24        &  -0.57&     $ 0.00^{+ 1.22}_{-0.00}$  \\
$D_{15}$(1675) & -6 &  19 $\pm$  8  & -9&   15 $\pm\ $9        &  -1.74&     $ 0.00^{+ 1.28}_{-0.00}$\\
$F_{15}$(1680) & 14 & -15 $\pm$  6  &142&  133 $\pm$ 12        &   0.44&     $ 0.00^{+ 1.18}_{-0.00}$\\
$F_{15}$       & -12 &                    &  5&                           &   0.78&                               \\
$F_{15}$(2000) & -1 &                    & 13&                           &  -0.38&                                  \\
$F_{17}$(1990) &  6 &   1                &  8&  4                        &  -1.25&     $ 0.00^{+ 2.17}_{-0.00}$ \\
\hline \hline\end{tabular} \label{Tab:AG}
\end{table*}

In Table~\ref{Tab:AG} we report on our results within the model $B$,
for all $n$~=1 and 2 shell resonances generated by the quark model
and complemented with the OGE model. In that Table, 2$^{nd}$ and
4$^{th}$ columns embody our results for the helicity amplitudes.
Those amplitudes are in lines with results from other similar
approaches (see Tables I and II in Ref.~\cite{Capstick2000}).

Comparing our results for the dominant known resonances of the model
$B$ with values reported in PDG~\cite{Yao2006} (3$^{rd}$ and
5$^{th}$ columns in Table~\ref{Tab:AG}) leads to following remarks:
{\it i)} $A_{1/2}$ amplitudes for $S_{11}$(1535) and $S_{11}$(1650),
as well as $A_{1/2}$ and $A_{3/2}$ for $D_{13}$(1520) and $A_{3/2}$
for $F_{15}$(1680) are in good agreement with the PDG values. For
this latter resonances the $A_{1/2}$ has the right magnitude, but
opposite sign with respect to the PDG value. However, for that
resonance $A_{3/2}$ being much larger than $A_{1/2}$, the effect of
this latter amplitude is not significant enough in computing the
observables. The amplitudes for $P_{13}$(1720) deviate significantly
from their PDG values, as it is the case in other relevant
approaches (see Table II in Ref.~\cite{Capstick2000}). Those large
values produced by our model forced us to leave the symmetry
breaking coefficient for $P_{13}$(1720) as a free parameter
(Table~\ref{Tab:GBE_1}) in order to suppress its otherwise too large
contribution. As much as other known resonances are concerned we get
results compatible with the PDG values for $D_{13}$(1700) and
$F_{17}$(1990), and to a less extent for $D_{15}$(1675). For
$P_{11}$(1440) our result deviates significantly from the PDG value.
Once again, our result confirms the general trend observed in other
works (see Table II in Ref.~\cite{Capstick2000}), which very likely
reflects the still unknown structure of that resonance. Finally, we
put forward predictions also for the missing resonances, for which
we find rather small amplitudes, explaining the negligible roles
played by them in our model.

The 6$^{th}$ and 7$^{th}$ columns in Table~\ref{Tab:AG} show our
results and PDG values, respectively, for the partial decay widths
of resonances decay in the $\eta N$ channel, where $\sigma$ is the
sign for $\pi~N\to~\eta~N$ as in Ref.~\cite{Koniuk1980}. Notice that
the sign ($\sigma$) in the PDG is known only for $S_{11}$(1535).
Except for the two star resonance $P_{13}$(1900), the theoretical
results are close to the PDG values.

It is worthwhile noticing that all dominated resonances in our model
$B$ have large helicity amplitudes, while some of them turn out to
have rather small decay widths to the $\eta N$ channel. This result
indicates that in looking for appropriate reactions to search for
missing resonances it is not enough to have rather sizeable decay
width, but one needs to put forward predictions for the observables.

%

\section{Summary and conclusions}\label{Sec:Conclu}

A formalism bringing together a chiral constituent quark approach
and one-gluon-exchange model was presented and used to derive
photoexcitation helicity amplitudes and partial decay width of the
nucleon resonances.

Our approach gives a reasonable account of the measured observables
for the process $\gamma~p~\to~\eta~p$ from threshold up to
$W~\approx$ 2 GeV. Among the twelve nucleon resonances in that
energy range, compiled by PDG, five of them are found to play
crucial roles in the reaction mechanism, namely, $S_{11}(1535)$,
$S_{11}(1650)$, $P_{13}(1720)$, $D_{13}(1520)$, and $F_{15}(1680)$.
However, those known resonances led to our model $A$, which does not
allow an acceptable description of the data. Five extra resonances
generated by the formalism, and known as missing resonances, turn
out to show no significant contributions to the process under
investigation. However, two new resonances reported in the
literature, $S_{11}$ and $D_{15}$, are found relevant to that
process; the most important effect comes from the $S_{11}$
resonance. We extracted the mass and width of those resonances:
$S_{11}$ [1.730 GeV, 217 MeV], and $D_{15}$ [2.090 GeV, 328 MeV].
Our model $B$, embodying those latter resonances, describes
successfully the data.

The helicity amplitudes and decay widths are calculated with the
same parameters. Our results are compatible with other findings and
come out close to the PDG values in most cases.

To go further, we are pursing our investigations in two directions,
\begin{itemize}

 \item {In the present work the {\it s-}channel resonances with
masses above 2 GeV were treated as degenerate, given that the
transition amplitudes, translated into the standard CGLN amplitudes
were restricted to harmonic oscillator shells $n \le 2$. recently,
we have extended our formalism and derived explicitly the amplitudes
also for $n$=~3~to~6 shells. Model search, including {\it all} known
one to four star resonances in PDG, for $W~\approx$ 2.6 GeV is in
progress~\cite{He2007}.}

 \item {Our constituent quark approach applied to the $\gamma~p~\to~K^+ \Lambda$
 channel~\cite{Julia2005}, showed that the intermediate meson-baryon states,
treated within a coupled channel formalism~\cite{Chiang2004}, have
significant effects on the photoproduction
observables~\cite{Julia2006}. A more sophisticated coupling-channel
treatment~\cite{Durand2008} has been developed and is being applied
to the $\eta$ photoproduction reaction. Results of that work will be
reported elsewhere.}
\end{itemize}
%
%
%

\section*{Acknowledgements}

We are deeply grateful to Qiang Zhao for enlightening discussions.

%


\begin{appendix}


%

\section{Mixing coefficients of the wave functions}
\label{apdx:ci}

In Table~\ref{Tab:GBE_masses}, we present the mixing coefficients of
the wave functions. In Ref.~\cite{Isgur1978,Isgur1979}, Isgur and
Karl have given their explicit values for positive parity and
negative parity resonances respectively. But in
Ref.~\cite{Isgur1978} the mixing between $n=0$ and $n=2$ shells is
not considered. Such mixings for the ground state are given in
Ref.~\cite{Koniuk1980} without the contribution of $^{2}P_A$. The
parameters in that reference are determined only by the mass
spectrum. Here we give our results by fitting both the mass spectrum
and the $\eta$ photoproduction observables. In calculation we follow
the conventions in Ref.~\cite{Koniuk1980}.

The mixing coefficients reported here lead to mixing angles, $\Theta
_S$ = -31.7$^\circ$ and $\Theta _D$ = 6.4$^\circ$ in agreement with
results from other
authors~\cite{IK,Chizma2003,Capstick2004,Bensafa}.

\begin{table*}[!hb]

\caption{Mixing coefficients of the wave functions. } \small

\renewcommand\tabcolsep{0.56cm}\begin{tabular}{ l | r rr rr rrrr }  \hline\hline

state&\multicolumn{4}{c}{wave function($^{2S+1}L_\pi$)}       \\
\hline\hline

 S11       &   $^{2}P_M$      &  $^{4}P_M$      &        &        &        \\

 N(1535)   & -0.851 &  0.526 &        &        &        \\

 N(1650)   &  0.526 &  0.851 &        &        &        \\\hline

 P11       &  $^{2}S_S$      & $^{4}D_M$        &  $^{2}P_A$       &  $^{2}S'_S$       &$^{2}S_M$    \\

 N(938)    &  0.941 & -0.043 & -0.002 & -0.260 & -0.211 \\

 N(1440)   &  0.268 &  0.000 &  0.000 &  0.964 &  0.006 \\

 N(1710)   &  0.175 & -0.343 & -0.071 & -0.054 &  0.919 \\

           & -0.103 & -0.839 & -0.424 &  0.031 & -0.324 \\

 N(2100)   & -0.032 & -0.421 &  0.903 &  0.010 & -0.080 \\ \hline

 P13       &  $^{2}D_S$      & $^{2}D_M$       &  $^{4}D_M$      &   $^{2}P_A$     &  $^{4}S_M$      \\

 N(1720)    &  0.858 & -0.483 &  0.023 & -0.003 & -0.176 \\

 N(1900)    &  0.314 &  0.234 & -0.365 &  0.095 &  0.839 \\

           & -0.185 & -0.482 &  0.606 & -0.333 &  0.505 \\

           &  0.359 &  0.686 &  0.496 & -0.387 & -0.065 \\

           & -0.059 & -0.096 & -0.502 & -0.854 & -0.073 \\ \hline

 D13       &  $^{2}P_M$      &  $^{4}P_M$   &        &        &        \\

 N(1520)   & -0.994 & -0.111 &        &        &        \\

 N(1700)   & -0.111 &  0.994 &        &        &        \\ \hline

 D15       &  $^{4}P_M$      &        &        &        &        \\

 N(1675)   &  1.000 &        &        &        &        \\ \hline

 F15       &   $^{2}D_S$  &  $^{2}D_M$      &  $^{4}D_M$      &        &        \\

 N(1680)   &  0.883 & -0.469 &  0.001 &        &        \\

           & -0.457 & -0.860 & -0.225 &        &        \\

 N(2000)  & -0.107 & -0.198 &  0.974 &        &        \\ \hline

 F17&  $^{4}D_M$     &        &        &        &        \\

 N(1990)   &  1.000 &        &        &        &        \\ \hline

 \hline\end{tabular}\label{Tab:GBE_masses}

\end{table*}


%
\newpage

\end{appendix}

%


%


%

\end{document}